# Terahertz Fano-like resonators based on free-standing metallic wire woven meshes

**DEJUN LIU,**[1,2,*] **AND TOSHIAKI HATTORI**[1,*]

[1]*Department of Applied Physics, University of Tsukuba, Tsukuba 305-8573, Japan*
[2]*Depatment of Physics, Shanghai Normal University, Shanghai 200234, China*
[*]*dejunliu1990@gmail.com*
[*]*hattori@bk.tsukuba.ac.jp*

**Abstract:** Most periodic terahertz (THz) structures need a substrate to support; thus, additional absorption occurs, resulting in a low quality (Q) factor. Free-standing structures that do not require any holder or substrate show high levels of flexibility and stretchability and hence are well-suited for THz applications. In this work, a free-standing THz metal structure consisting of metallic wire woven meshes is proposed and demonstrated. Experimental and numerical results exhibit that this metallic mesh achieves a sharp Fano-like resonance dip, which has not been found in previous studies. Investigation results indicate that the high Q Fano-like resonance dip comes from the single-layer metal bent wire because of its bending effect. The resonance field longitudinally covers the input and output end faces due to the large field volume of the woven meshes and benefits from near-field sensing applications.

## 1. Introduction

Terahertz (THz) waves (0.1–10 THz) have been explored and applied to wide practical applications, such as imaging, sensing, and spectroscopy, due to their unique properties, namely, non-ionization and low photon energy (4.1 meV at 1 THz) [1-2]. For example, many molecules show the vibrational and rotational polar modes in THz regions, making THz spectroscopy an ideal tool for biological sensing [2]. However, most THz systems are based on bulky free-space optics, which needs precise alignment and servicing; thus, the application of THz systems is limited [3]. A mismatch between the THz waves and the sensed target size exists because the use of THz waves is hindered by its long wavelength in relation to the size of the sample to be characterized [4].

    Plasmonic structures can overcome some of these challenges because of their enhanced field and high confinement [5]. Plasmonic structures with surface plasmon polaritons (SPPs) based on periodic structures have been investigated in the past decades [6-12]. Several THz plasmonic structures with different surface metallic patterns have been recently demonstrated; in these structures, strongly localized fields are concentrated on the metal surface [13-22]. Typical structures, such as metal groove arrays [23], metal gratings [24], and rectangular blind holes [25], have been reported. These structures are limited to THz sensing applications because they need a bulky prism or another additional coupler to excite SPPs. Plasmonic metamaterials can be applied to THz sensing because no additional coupler is needed. The metamaterials are composed of subwavelength periodic metallic resonators in which the localized fields are strongly enhanced. Metamaterials, such as asymmetric split ring resonators [19-20], C-shaped resonators [21], and corrugated metallic disks [22], have been proposed. Sharp Fano resonances with high quality (Q) factors can be obtained by breaking the structural symmetry of metamaterials. These metamaterials need dielectric substrates to sustain, resulting in additional material absorptions. Free-standing structures do not need additional substrates and couplers offer wide potential applications in THz regions. Free-standing metal hole arrays (MHAs) with Fano resonance under oblique wave incidences are used for DNA molecule detection [16-18]. Double-layer MHAs have been investigated to enhance the SPP field. The enhanced localized

SPP field is confined inside the gap between two layers [26, 27]. However, the enhanced field is difficult to manipulate, and double-layer MHAs suffer from high losses. Two-layered MHA structures are noncompact and difficult to control precisely. Thus, a mechanically self-supported metal structure with a miniature size is highly required. The self-supporting structure of woven-steel meshes has been investigated due to its properties, such as flexible design, deformability, and potentially low cost [28-29]. Abnormal group velocities with 88% high-power transmission can be achieved by the woven-steel mesh in the sub-1 THz regime. The THz spectroscopic analysis on the meshes is revealed, but the sharp spectral and modal properties induced by the mesh structure remain unknown.

In this work, a free-standing metal structure based on metallic wire woven meshes (MWWM) with sharp resonance dips is experimentally and numerically investigated [30]. The structural unit of the woven metal wire is critical to perform sharp Fano-like resonance modes due to its asymmetry. The double-layer woven mesh is simplified to single-layer metallic bent wire arrays to understand the origins of sharp resonance dips further. Numerical results demonstrate that the sharp dip originates from the bending effect in the metallic bent-wire arrays. The resonance frequency, bandwidth, and Q-factor of sharp dips can be modulated by changing the bending parameter of metallic bent wire arrays. High Q-factor resonance is optimized using the dimensions of the bending section of the wire and achieved to confine a large volume of resonance field. The resonance field longitudinally covers the input and output end faces due to the large field volume of the woven wires and benefits from near-field sensing applications.

## 2. Scheme of MWWM

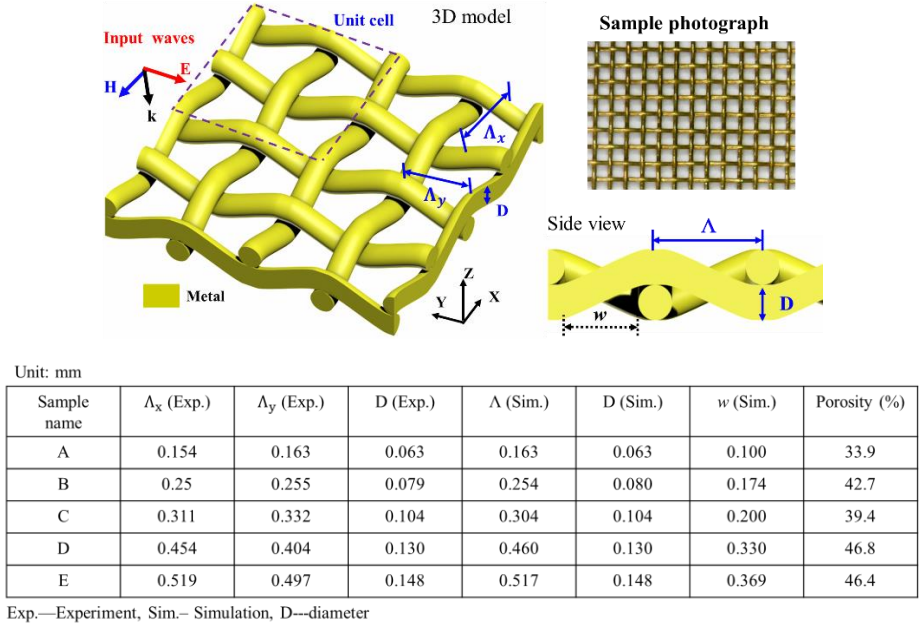

Unit: mm

| Sample name | $\Lambda_x$ (Exp.) | $\Lambda_y$ (Exp.) | D (Exp.) | $\Lambda$ (Sim.) | D (Sim.) | w (Sim.) | Porosity (%) |
|---|---|---|---|---|---|---|---|
| A | 0.154 | 0.163 | 0.063 | 0.163 | 0.063 | 0.100 | 33.9 |
| B | 0.25 | 0.255 | 0.079 | 0.254 | 0.080 | 0.174 | 42.7 |
| C | 0.311 | 0.332 | 0.104 | 0.304 | 0.104 | 0.200 | 39.4 |
| D | 0.454 | 0.404 | 0.130 | 0.460 | 0.130 | 0.330 | 46.8 |
| E | 0.519 | 0.497 | 0.148 | 0.517 | 0.148 | 0.369 | 46.4 |

Exp.—Experiment, Sim.— Simulation, D---diameter

Fig. 1. Configuration of 3D structures of MWWM and experiment sample photography and table with the structural parameters of woven meshes.

Figure 1 schematically plots the portion (2×2 cells) of the 3D structure of MWWMs with the square holes constructed by vertical cross layers of periodic metal wires. The experimental samples are either made of copper or nickel. For instance, the material of sample B is nickel, and that of the other samples is copper. Figure 1 exhibits a microscopic photograph. In simulation, the cells along the X- and Y-axes are infinitely and periodically extended. The perfectly matched layers are occupied along the Z-axis of the MWWM. During FDTD simulation, the material of woven meshes is assumed to be perfect electric conductors. A plane wave, which is approximate to the parallel beam with a size of 10 mm, is used in FDTD. The electric fields of the input transverse magnetic (TM) and transverse electric (TE) waves are perpendicular to the Y- and X-axes, respectively. The fraction of open area for sample B is 42.7%, which is calculated by $T_{porosity} = \frac{\pi}{2\sqrt{3}}(\frac{w}{\Lambda})^2$ [31], where $w$ and $\Lambda$ are the hole size and lattice constant, respectively [28]. In contrast to thin-film metamaterials [19-22], the MWWM structure is mechanically self-supporting without using a dielectric substrate; thus, it opens important degrees of freedom to the design of multilayer stacked metamaterials, waveguides, and antennas. The above-mentioned table shows the measured and simulated structural parameters. The sample's parameters in the X and Y directions have slightly different ($\Lambda_x \neq \Lambda_y$), indicating that the shape of the mesh holes is not square. $\Lambda_x = \Lambda_y = \Lambda$ is set in 3D-FDTD simulation to simplify the simulation model.

## 3. Results and discussion

### 3.1 Experimental and simulated spectra of MWWMs

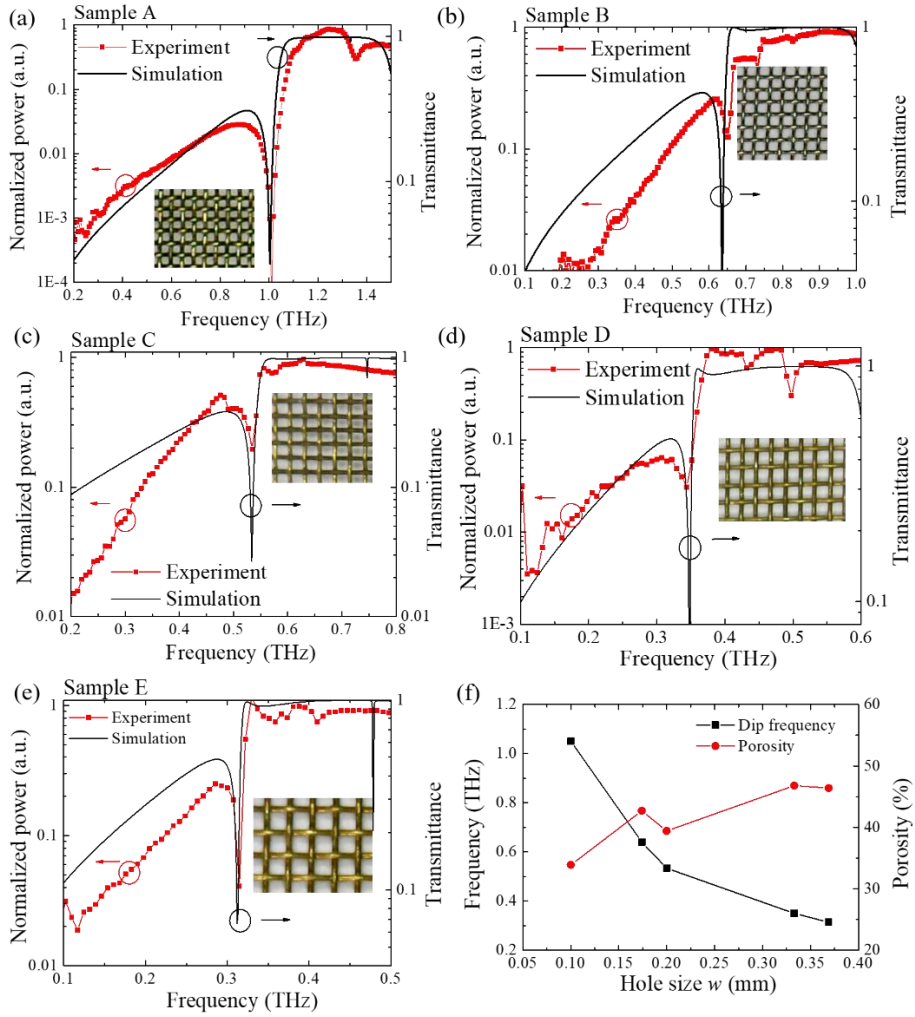

Fig. 2. (a–e) Experimental and simulated transmission spectra of MWMMs with various structural parameters.

The experimental results measured by the THz-TDS are shown in Fig. 2, where the simulation results calculated by FDTD are in good agreement. The spectral power of the experimental results is normalized to compare the spectral performance of meshes. Figs. 2 (a–e) illustrate that clear sharp dips with distinctly low transmittance can be observed in the transmission spectra. In a previous work with the same meshes [28], sharp dips were not found in these transmission spectra probably because of the THz beam size in the experiment. For example, a 10 mm parallel beam is used for this measurement, whereas a focused one with a smaller size was utilized in the previous work. The measured dip frequencies are 1.05, 0.63, 0.534, 0.35, and 0.314 THz for the 0.163, 0.254, 0.304, 0.460, and 0.517 mm-Λ meshes, respectively. The small discrepancy between the experimental and numerical results comes from the asymmetric holes in meshes and rough samples. Fig. 2 (f) demonstrates the relation between the hole width and the sharp dip frequency and porosity.

The sharp dip shifts to low frequencies as the hole width increases. This phenomenon suggests that the resonance dip is sensitive to structural parameter changes. In Fig. 3, we summarize the results of the sharp dip wavelength and structural parameters, such as the lattice constant (Λ) and hole width (w), to have an in-depth understanding of the relation between structural parameters and sharp dips. The dip wavelength shifts to a long one as Λ and w increase. Fig. 3 (a) shows the linear fit line of the dip wavelength and lattice constant. The dip wavelength shifts to long ones with the increasing mesh lattice constant. The data are well-fitted to the linear equation $\lambda_{dip} = 1.848 * \Lambda$, where the adjusted coefficient of determination ($R^2$) is 0.999. Fig. 3 (b) presents the relation between the dip wavelength and the hole width (w). The increasing hole width causes the dip wavelength to shift to long ones. The equation $\lambda_{dip} = 2.629 * w$ is used to fit the data, where $R^2$ is 0.992. The trend of the resonance wavelength shifts is consistent with that of the MHA investigated in previous studies wherein the mesh hole width plays a critical role [15-16].

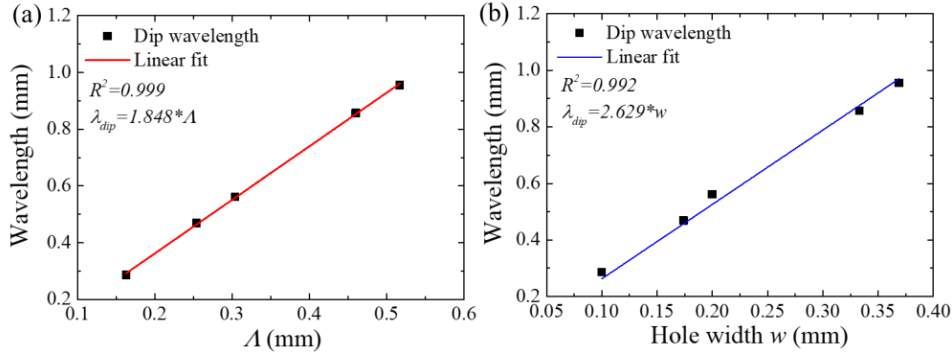

Fig. 3. Shifts of the dip wavelength with the change of the lattice constant (Λ) (a) and hole width (w) (b).

*3.2 Sharp resonance dips dependent on the structural parameters*

The 3D-FDTD method is used to compare the transmittance spectra between the MWWM and the planar MHA with the same metal wire diameter (D) and hole width of 0.08 and 0.174 mm, respectively. This task is performed to study the induced sharp dips further. Fig. 4 presents the results. In the MHA, the transmittance peak is located at 1.0 THz. The resonance wavelength is approximately 1.73 times the structural hole width (0.174 mm), corresponding to the lowest mode of SPPs [15, 18]. The transmittance of the spectral peaks is correlated to the field resonance in the X- or Y-axes inside the hole. The frequency of the surface-confined field on the MHA is mainly determined by the hole width, and its SPP field can be used in sensing applications [15-16, 18].

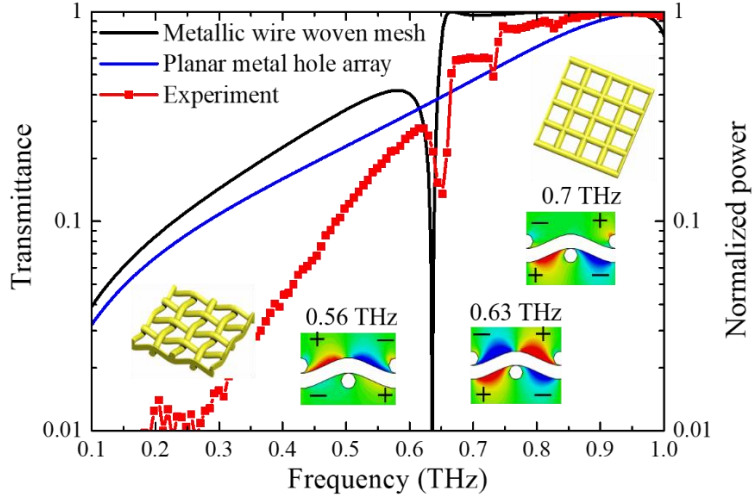

Fig. 4. Transmittance spectra of MWWM and planar MHAs. (inset) THz electric field ($E_z$) distributions in the X-Z plane at 0.56, 0.63, and 0.7 THz, respectively.

In the MWWM, one clear sharp dip with a high Q of 47 can be observed at 0.63 THz in the transmission spectrum. This sharp dip with nearby blocking propagation can be excited by TM and TE modes resulting from the metallic bending wires. Fano resonance indicates that a discrete bound state is coupled with a continued state [32-37]. The profile of this sharp dip resembles that of Fano resonance modes and is thus termed as Fano-like resonances. The insets in Fig. 4 illustrate the Z-axial distribution ($E_z$) of the electric fields at 0.56, 0.63, and 0.70 THz, representing the waves at the low resonance peak, resonance spectral dip, and high transmission band, respectively. The field patterns on the metal–air interfaces exhibit opposite dipolar plasmonic modes, which are similar to those of MHAs. The high intensity of the electric field is strongly confined in the wire edge and gradually decays along the Z-direction away from the mesh–air interface, and it can be regarded as SPP fields. The fields of low/high transmission band are located only at the input/output end face. However, the resonance field at 0.63 THz is located at both end faces. This situation results in a high power–distinction ratio with a spectral peak of approximately 1000. The wire diameter is a critical factor to the spectral dip in this MWWM structure. The field coupling effect between the top and bottom faces of the metal wire can be controlled by changing the wire diameter. The high confinement of SPP modes associated with the sharp Fano-like dips makes the MWWM suitable as THz sensors and filters.

The strong coupling in the metal structures results in evident phase $\varphi$ changes. Fig. 5 (a) shows the phase of air space, MWWMs, and planar MHAs. A sharp phase change at 0.63 THz can be observed in MWWMs. The MWWM shows a $0.47\pi$ phase shift ($\Delta\varphi$) in comparison with that of planar MHAs. The phase of MWWM is larger than that of air space in a frequency range of 0.1–1.0 THz. We calculate the effective refractive index of MWWM using $n=(\lambda\varphi/2\pi t)+1$, where $t$ is the structural thickness. As shown in Fig. 5 (b), a sharp dip occurs at a resonance frequency of 0.63 THz in the effective refractive index spectrum. In the frequency range of 0.1–0.95 THz, the effective refractive index of MWWMS is lower than $n=1$, indicating that the phase velocity of the light in the MWWM is faster than the speed of light; this finding is in good agreement with reference [28]. Therefore, the resonance dip originates from the strong coupling between the EM waves and the MWWMs, resulting in a sharp change in the phase, transmission spectrum, and effective refractive index.

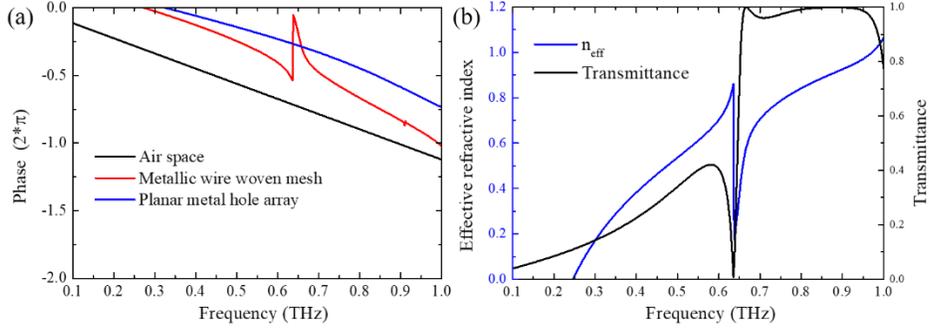

Fig. 5. Phase (a) and effective refractive index (b) of MWWMs.

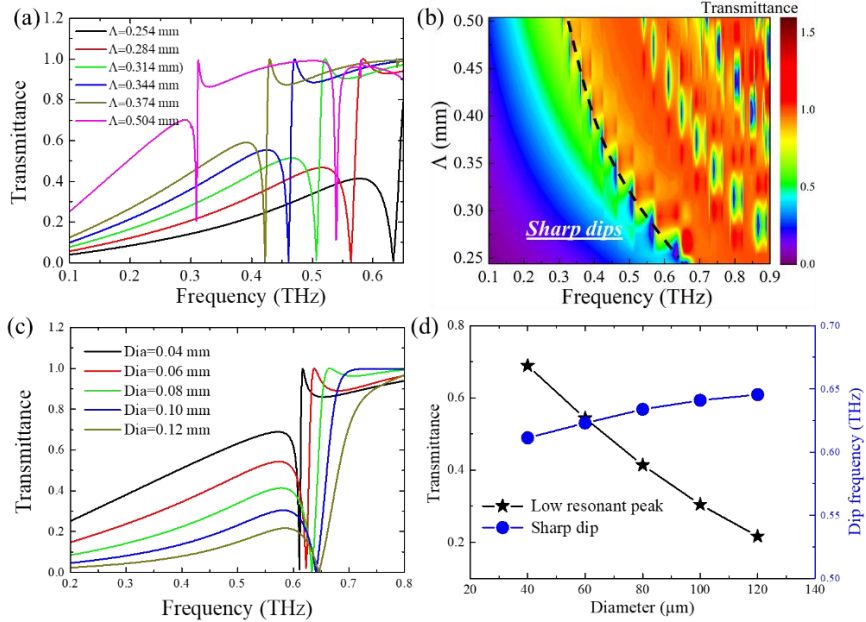

Fig. 6. Transmission spectra of MWWMs with various lattice constants (a) and the contour map (b). Transmission spectra of MWWMs with various diameters (c) and the relation among the wire diameter, resonance dip frequency, and low resonance peak transmittance (d).

The transmission spectra with various mesh lattice constants and wire diameters are calculated and depicted in Fig. 6 to characterize the dependence of the structural parameters on spectral dips. The resonance dip shifts to low frequencies with the increase in the lattice constant, whereas the other parameters maintain fixed values (Fig. 6 (a)). The resonance dip wavelength is dominated by the hole width; the equation can be expressed as $\lambda \sim 2nw$, where $n$ and $w$ are the refractive index and hole width, respectively [36-37]. The linewidth of the sharp resonance dips shortens with the increase in the lattice constant. The frequency and bandwidth of the resonance dips can be critically controlled by the lattice constant of MWWMs. Fig. 6 (b) presents the contour map of the transmission spectra for MWWMs with various Λ. The results demonstrate that the large lattice constant results in low resonance frequencies. The wire diameter of MWWMs is a critical parameter to manipulate the resonance dip. The wire diameter $D$ is varied, whereas the lattice constant is fixed at 0.254 mm to study the wire diameter dependence of the geometrical structures on the dip. Figs. 6 (c–d) present the results. When the diameter of the metallic wire is 0.04 mm, the resonance dip achieves a Q-factor of 72. With the wire diameter increase from 0.04 mm to 0.12 mm, the dip linewidth becomes large, and the Q-factor reduces from 72 to 23. The transmittance of the low resonance peak is also dissipated with the increased wire diameter. When the wire diameter is sufficiently large, the SPPs on both sides show a weak coupling effect. The reduced transmitted magnitude comes from the low coupling efficiency of SPPs between the input and the output faces of MWWMs. The wire diameter considerably affects the resonance dip and high passband because the diameter alteration changes the effective plasma frequency of MWWMs [28, 38].

*3.3 Sharp resonance dips excited by single-layer metallic bent wire arrays*

Fig. 4 shows that a strong resonance occurs on the transmittance spectra of MHA when the periodic hole array is constructed by two across layers of the periodic metal wires with a 3D structure. We analyze the spectral characteristics and resonance field by simplifying the woven mesh to a single-layer metallic bent wire array to understand the mechanism of induced Fano-like resonance in MWWMs efficiently. Lattice constant ($\Lambda$) of the metallic wire array is 0.254 mm, and the wire diameter ($D$) is 0.08 mm. The single-layer metallic wire array is a straight wire array. The metallic bent wire arrays consist of periodic bent wires positioned in-phase and out-phase. The TM and TE modes are perpendicular and parallel to the wire array, respectively. The in- and out-phases of the bent wire arrays are presented in Figs. 7 and 8.

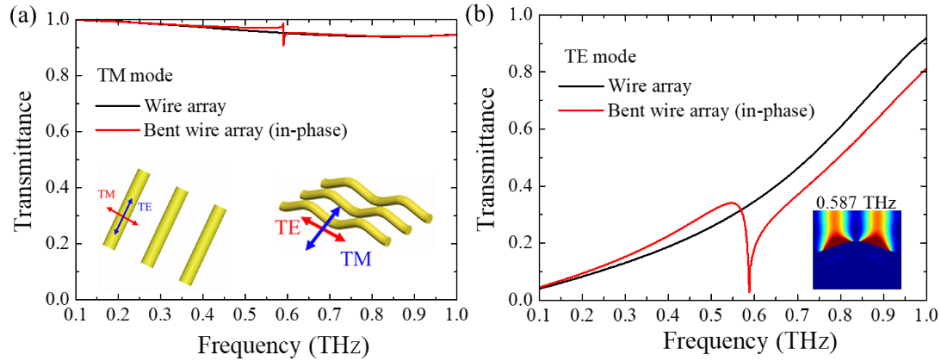

Fig. 7. Transmission spectra of the single-layer metallic wire and metallic bent wire (in-phase) arrays for the TM (a) and TE (b) modes. The inset shows the electric field distribution of the TE modes at 0.587 THz.

Figs. 7 (a) and (b) depict the calculation transmission spectra of the single-layer metallic wire and bent wire arrays. The bent wire array's arrangement is in-phase. The metallic wire arrays with a subwavelength slit have been studied in a previous work [30]. Under the TM modes, the electric field is perpendicular to the wire array, and a small discrepancy can be found in the transmission spectra of the metallic and bent wire arrays (Fig. 7 (a)). A sharp dip with low transmittance occurs at 0.587 THz when the TE mode incidents into the bent wire array (Fig. 7 (b)). A sharp dip can be induced by the TE mode rather than the TM one. The TE mode electric field is parallel to the bent wire array. The induced Fano-like resonance dip comes from the continuum spectrum constructively or destructively interacting with the bending wire through an interference effect. The inset in Fig. 7 (b) exhibits the electric field distribution; a strong localized field is concentrated at the bending regions of the bent wire arrays. The field exponentially decays away from the metal surface and can be regarded as an SPP field.

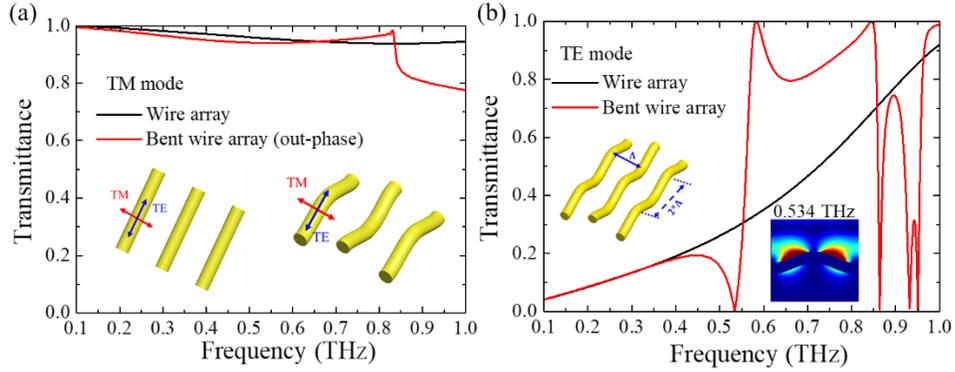

Fig. 8. Transmission spectra of the single-layer metallic wire and metallic bent wire (out-phase) arrays for the TM (a) and TE (b) modes. The inset exhibits the electric field distribution of the TE modes at 0.534 THz.

A similar phenomenon can be found in Fig. 8. Fig. 8 (a) presents that the sharp dip under the TM modes has not occurred in the transmission spectrum of the bent wire arrays. In contrast to the straight wire arrays, the out-phase bent wire array under TE modes exhibits a sharp Fano-like dip with nearly zero transmittance at 0.534 THz. The high pass-band has been extended from 0.59 THz to 0.85 THz. This result suggests that the out-phase alignment makes the structure compact. The bending effect accelerates the mesh application with compact size in 3D space. This bending effect can also be used for 2D metasurface designing. The inset in Fig. 8 (b) is the electric field distribution. The out-phase bent wire array confined at the 0.534 THz field in the bending regions is similar to that of in-phase ones. The tail of the induced field in the out-phase wire array is shorter than that of the in-phase one (inset in Fig. 7 (b)) due to the out-phase arrangement.

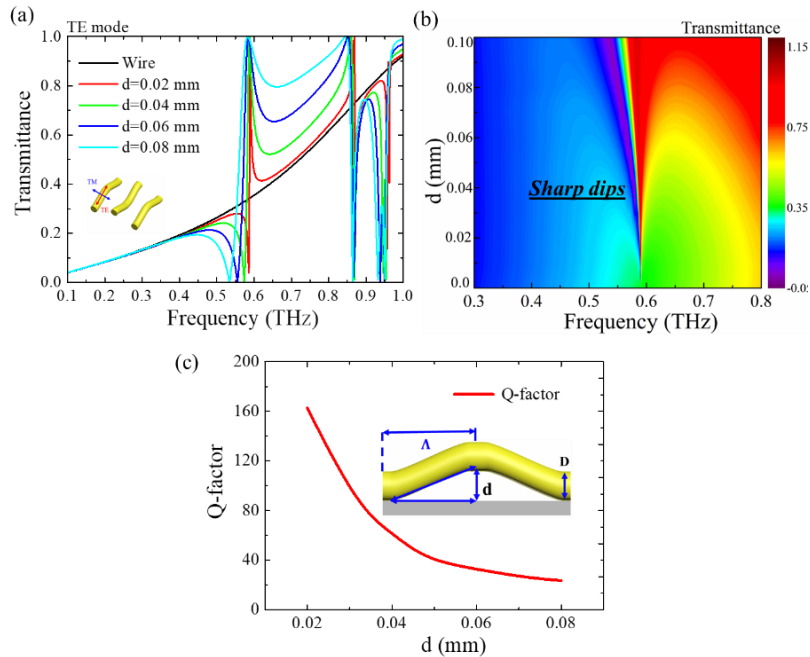

Fig. 9. Spectra of the single-layer metallic bent wire (out-phase) (a). Contour map of the transmittance spectra for the single-layer bent wire arrays with various bending parameters (b). Q-factor of the resonance dips of bent wire arrays (c).

In this section, we analyze the response of the Fano-like resonance dips by changing the bending parameters in the single-layer metallic bent wire arrays (out-phase) (Fig. 9). The inset in Fig. 9 (c) illustrates that the Z-axial space above the X–Y plane surface can be defined as the bending parameter (*d*). In Fig. 9 (a), the transmission spectra of the bent wire arrays with a bending parameter of 0.08 mm achieve one sharp dip at 0.534 THz in comparison with metal wire arrays without bending. With the bending parameter increase from 0.02 mm to 0.08 mm, the Fano-like resonance frequency shows a redshift of $\Delta f$=0.051 THz. Fig. 9 (b) shows the contour map of the transmittance spectra for the single-layer bent wire arrays with different bending parameters. One clear sharp dip is induced when the bending parameter of the metallic wire arrays is larger than zero. The bandwidth of the resonance mode decreases with the decrease in bending parameters in correspondence to the enhancement of the resonance Q factor. The performance trend of the Q factor versus the bending-factor variation is summarized in Fig. 9 (c). The Q-factor can be calculated by $f_0/\Delta f$, where $f_0$ and $f$ is the resonance frequency at the dip and the full-width at half-maximum ($FWHM = \Delta f = f_2 - f_1$) bandwidth, respectively. The investigation shows that the maximum Q-factor of approximately 162.6 can be achieved when the bending parameter is 0.02 mm. The sharp dip Q-factor can be improved by reducing the bending parameters of the metallic wire arrays. The bending parameter in the metallic wire arrays are critical to induce the spectral Fano-like resonance dip and different bent levels formed various 3D structures can modulate the spectral range of the field resonance.

### 3.4 Modal field distribution of MWWMs

As discussed in the previous sections, the coupling between the EM waves and the metallic structures results in an enhanced field on the metal–air interface. We explore the modal field distribution of MWWMs in this section. We simulate the electric field distribution at 0.56, 0.63, and 0.70 THz in the X–Y (Z=0 mm) and Z–X cut plane (Y=0 mm), representing the waves at the low resonance peak, sharp Fano-like dip, and high transmission band, respectively. Fig. 10 depicts the results.

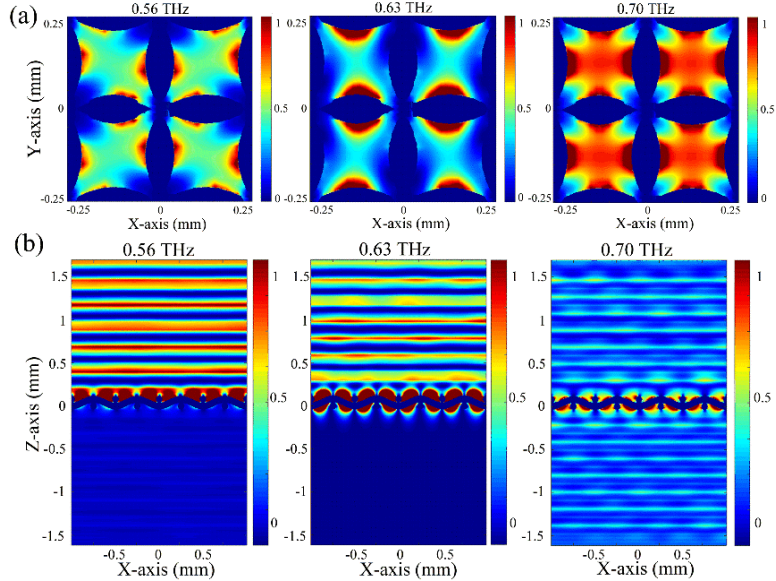

Fig. 10. Electric field distribution in the X–Y (a) and Z–X (b) cut planes at 0.56, 0.63, and 0.70 THz.

At 0.56 THz, a strong SPP field is located at the mesh edge in the X and Y directions. In the resonance dip at 0.63 THz, the localized electric field only distributes in the Y direction. The symmetrical field distribution in the X and Y directions can be found at 0.7 THz. The induced field is strongly confined inside the mesh cavities. Fig. 10 (b) shows the electric field distribution of these frequencies in the Z–X cut plane. A small part of the 0.56 THz field passes through the MWWMs. In the resonance dip at 0.63 THz, the THz waves are completely blocked by the meshes. Strong electric fields cover the input and output end faces of the woven wires, which benefits THz sensing applications. An extraordinary optical transmission occurs at a high frequency of 0.70 THz, and the electric fields are primarily confined inside the mesh cavities (Figs. 10 (a) and (b)).

The plasmonic structures show huge potential in various applications, such as near-field imaging and sensing, because of the enhanced localized SPP field [39-40]. The MWWM is considered a plasmonic structure, which can be used for THz sensing. In this section, the near-field distribution of MWWMs at frequencies of 0.56, 0.63, and 0.70 THz is investigated. Fig. 11 presents the near-field distribution of MWWM at 0.56, 0.63, and 0.70 THz. The gray area is the metal wire region. The modal patterns at various frequencies demonstrate that the maximum fields are located at the metal wire edges because of the evanescent modes. The field enhancement factor is defined as the ratio of the metal surface and incident fields [41-42]. At 0.56 THz, the enhanced field is confined at the top face of the metal wires, and the enhancement factor is approximately three. It indicates that the SPP field on the metal surface is three times larger than the incident fields. A maximum field enhancement factor of 8.5 can be achieved at 0.63 THz. The field decay length ($1/e$) is approximately 0.04 mm, which corresponds to $\lambda/12$, achieving a high accuracy for THz near-field imaging [39]. The electric field at 0.70 THz shows an enhancement factor of 2.8, which is lower than that of 0.56 and 0.63 THz.

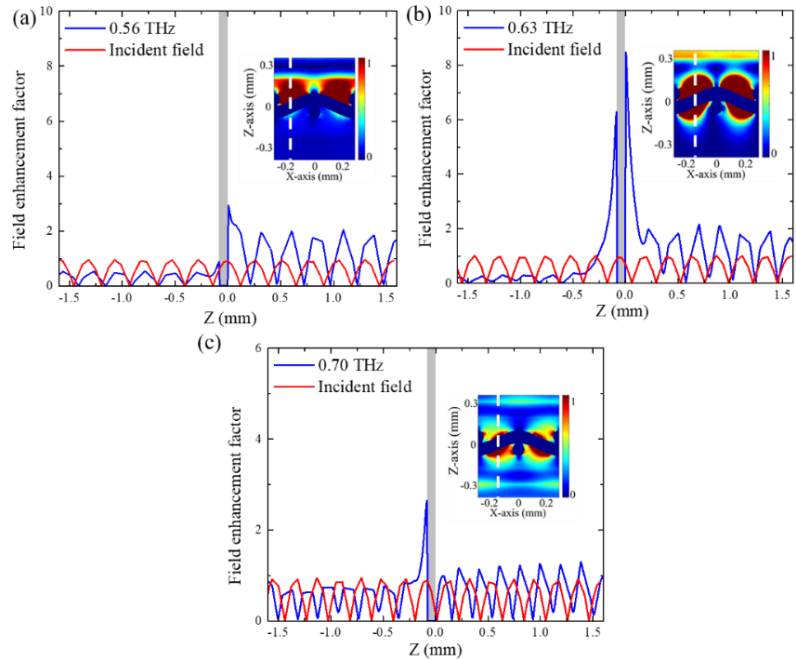

Fig. 11. Near electric field distribution of MWWMs at 0.56 (a), 0.63 (b), and 0.70 THz (c). The inset shows one cell of MWWMs, and the gray area represents the metal wires.

Fig. 12 shows the power density distribution and energy flow at the center of MWWMs. Frequencies 0.56, 0.63, and 0.70 THz are selected as examples to investigate. A high energy density in the cavity of meshes is achieved at 0.70 THz. A loose power density can be observed at 0.56 and 0.63 THz. Fig. 12 (b) presents the power vector distributions. A phenomenon of circumfluence-like energy flows can be found in the mesh cavities, comes from the bending effect within meshes. Interestingly, the flow direction is out-phase in the neighboring cavities. Besides, the energy flow patterns at 0.70 THz are opposite to that of 0.56 and 0.63 THz.

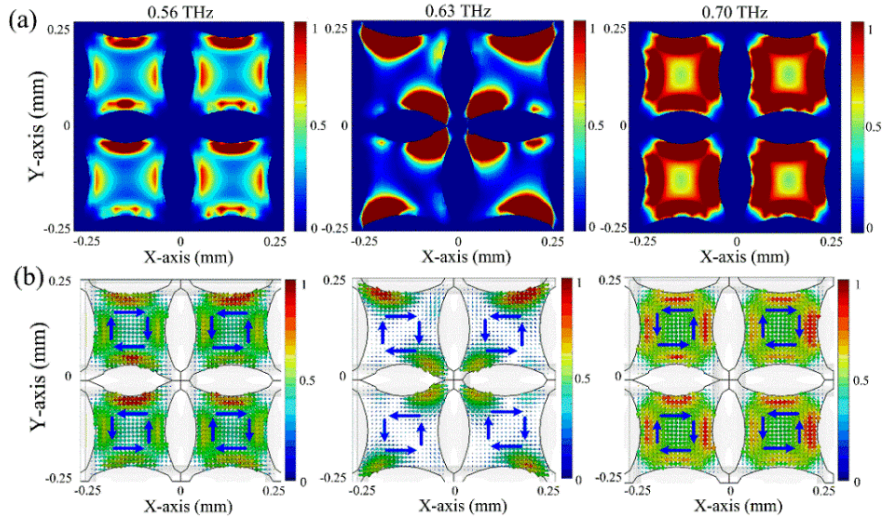

Fig. 12. Power density distribution (a) and power vector (b) of MWWM at 0.56, 0.63, and 0.70 THz.

## 4. Conclusion

Free-standing MWWMs in THz frequencies are investigated. The investigation results demonstrate that a sharp Fano-like resonance dip with a high power–distinction ratio can be induced by the MWWM. The woven mesh is divided into single-layer metallic bent wire arrays to study the origin of sharp resonance dips. A sharp resonance dip with low transmittance is induced when the incidence waves are TE modes because of the bending effect within bent wire arrays. The field enhancement, modal confinement, and Q-factor can be manipulated by tuning the bending parameter. The Fano-like resonance dip at 0.63 THz for 0.254 mm-Λ MWWM shows a high Q-factor of 47, obtaining a field enhancement factor of 8.5. The maximum field decay length ($1/e$) is approximately 0.04 mm, corresponding to $\lambda/12$. The resonance field longitudinally covers the input and output end faces of the woven mesh, thereby obtaining a large field volume. The investigation result reveals that MWWM improves the surface field of MHA for THz wave sensing applications. Therefore, MWWMs are suitable as THz sensing devices because of the sharp spectral and free-standing properties.

## Funding

This study received funding from the China Scholarship Council (CSC NO. 201606890003).

## Acknowledgments

The authors would like to thank KGS (www.kgsupport.com) for providing an English language review.

## Author contributions

D. Liu performed the measurements and simulations at the University of Tsukuba. D. Liu analyzed data and wrote the paper. T. Hattori provided measurement samples.

## Disclosures

The authors declare no conflicts of interest.